\documentclass[11pt,a4paper]{article}
\usepackage{jheppub}

\title{The Hagedorn spectrum and large $N_c$ QCD in 2+1 and 3+1 dimensions}

\author[a]{Thomas D. Cohen}
\author[a]{and Vojt\v{e}ch Krej\v{c}i\v{r}\'{i}k}

\affiliation[a]{Maryland Center for Fundamental Physics\\
Department of Physics, University of Maryland, College Park, MD 20742-4111}

\emailAdd{cohen@physics.umd.edu}
\emailAdd{vkrejcir@umd.edu}

\abstract{
We show that a Hagedorn spectrum (i.e., spectrum where the number of hadrons grows exponentially with the mass) emerges automatically in large $N_c$ QCD in 2+1 and 3+1 dimensions.   The approach is based on the study of Euclidean space correlation functions for composite operators constructed from quark and gluon fields and exploits the fact that the short time behavior of the correlators is known in QCD.  The demonstration relies on one critical assumption: that perturbation theory accurately describes the trace of the  logarithm of a matrix of  point-to-point correlation functions in the regime where the perturbative  corrections to the asymptotically free value are small.
}


\arxivnumber{--}

\begin{document}
\maketitle

\section{Introduction}

One of the oldest questions in strong interaction physics is the density of hadrons in the spectrum as a function of mass for large mass.
It was conjectured long ago by Hagedorn that this density (when suitably averaged)  grew exponentially with
the mass \cite{Hagedorn1, Hagedorn2}.  A useful way to parameterize this is via its integral,  $N(m)$, the number
of hadrons with mass less than $m$.  One way to state Hagedorn's conjecture is that at asymptotically large $m$,
\begin{equation}
N(m) \sim \left ( \frac{m}{T_H}\right )^a \exp \left(\frac{m}{T_H}\right) \; ,
\end{equation}
where $T_H$, the so-called Hagedorn temperature, is a parameter controlling the exponential growth.
Note that in a simple model, where hadrons are treated as a noninteracting free gas, the Hagedorn temperature
represents an upper bound on the temperature of a hadronic phase of matter as the energy density diverges for $T>T_H$.
The power-law prefactor plays an important role in attempts to fit the Hagedorn spectrum from
data \cite{bronflorgloz}   and also determines the thermodynamic behavior of strongly interacting matter
as $T_H$ is approached from below in the simple noninteracting hadron gas model \cite{Cohen2006}.
A more useful way to state Hagedorn's conjecture for the purposes of this paper is that for asymptotically
large masses there exists positive value of  $T$ such that
\begin{equation}
N(m) \geq e^{m/T} \;  ;
\label{hagedorn}
\end{equation}
$T_H$ is the maximum value of $T$ for which eq.~(\ref{hagedorn}) holds.

\begin{figure}
\begin{center}

\includegraphics[width=10cm]{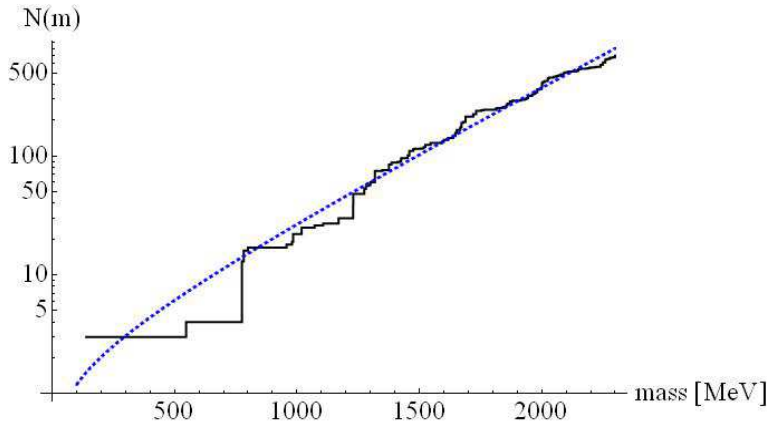}

\caption{$N(m)$ for nonstrange mesons using mesons masses extracted from various hadronic processes extracted reported by the Particle Data Group~\cite{PDG}.  The fit is of the form $N(m)=am^be^{m/T_H}$  and  yields a Hagedorn temperature of 426~MeV.  The fits were
done for mesons with masses up to 2300~MeV. Our estimate of the Hagedorn temperature is consistent
with the results of ref.~\cite{bronflorgloz}. Note that it is almost 2.5 times larger than
the critical temperature $T_C$ obtained by lattice gauge calculations \cite{ChengChrist}.
}
\label{empirical}

\end{center}
\end{figure}

As shown in figure \ref{empirical}, the extracted masses
of hadronic resonances \cite{PDG}, $N(m)$ does, indeed, grow very rapidly up to the
point where it becomes difficult to extract resonance parameters from experimental data (around 2 GeV).  This behavior appears to be consistent with the notion that QCD does have a Hagedorn spectrum.
However, it is very difficult to establish Hagedorn's conjecture in a compelling way from the empirical data.
In part this is a practical issue; one would need to extract hadron masses for hadrons up to much larger masses to get
compelling evidence for an exponential growth.  Moreover, underlying this practical issue is an important theoretical
one: highly excited hadrons are not particles; they are resonances and as such do not have well-defined masses.
The mass parameters can only be extracted from partial wave analysis of various scattering  processes using some
model dependent assumptions.  Such model-dependence is quite weak for well-isolated narrow resonances and for these one
can state masses with some level of confidence.  However as resonances in some channel become wide  or close to each other,
such model dependence grows and it becomes difficult to isolate resonant state in a meaningful way.
Moreover, \emph{any} model dependence in the meaning of a hadron's mass makes the issue of the density of hadronic states intellectually problematic.

Before attempting to deal with the problem of ill-defined hadron masses, it is useful to understand why one might expect QCD to have a Hagedorn spectrum.
 Recall that Hagedorn spectra arises automatically in simple string theories
\cite{Strings} with unbreakable and noninteracting strings.  It is noteworthy that string theory  was originally formulated as a theory of strong interaction.  Moreover, given confinement it is plausible that highly excited states in QCD should act stringy.  For the case of pure gauge theory there  is strong evidence \cite{Lattice}  that   widely separated  static quark sources have a linearly rising  potential, (i.e., an area law for the Wilson loop).  This arises because for widely separated sources, the flux arranges itself into tubes with a characteristic width and fixed energy per
unit length \cite{Lattice}.  It is plausible that for highly excited states,
which would be expected to have flux tubes which are much longer than their
width would act dynamically as strings  and as an effective string theory  would naturally give rise to a Hagedorn spectrum.  Mesons in such a picture are interpreted as open strings.   However, this picture is flawed.  It is based on pure gauge theory, in which confinement implies unbreakable strings.  In QCD, with dynamical quarks, flux tubes can break.  Indeed, this is the same issue as noted above --- the fact that flux tubes break implies that mesons decay and thus can only be seen as resonances with nonzero widths.

It is not clear whether there is a clean way to deal with this issue in an unambiguous way for QCD in the physical world.  However, if one focuses on the large $N_c$ limit of QCD \cite{tHooft, Witten}, the issue vanishes.  As the large $N_c$ limit is approached, meson decays are suppressed by a factor of $1/N_c$; flux tubes do not break and mesons become stable.  The goal of the present work is to show that at large $N_c$ QCD must have a phase transition.  This is at least a well-posed theoretical question.  Of course, the question of whether or not this is of phenomenological relevance depends on how close the $N_c=3$ world is to the large $N_c$ world.

The first  derivation  of a Hagedorn spectrum  in some variant of  large $N_c$ QCD was done
by Kogan and Zhitnitsky \cite{KoganZhitnitsky}, who explicitly  computed the spectrum of
 large $N_c$ QCD in 1+1 dimensions  with adjoint
fermions and showed that it possess a Hagedorn-type behavior.    Ideally one could similarly compute the spectrum for large $N_c$  QCD for more than one spatial dimension.  However, in practice we do not know how to solve for the spectrum.  Numerical studies using lattice QCD are poorly suited for extracting high-lying stars.  There was a study of the large $N_c$ glueball spectrum based on a numerical treatment of a transverse lattice QCD in a light cone formalism
\cite{DalleydeSande}; while the results are consistent with the Hagedorn spectrum, the evidence was not definitive.  There are, however, indirect ways to probe the issue.   One is by the study of QCD thermodynamics.  It is well known that large $N_c$ QCD has a first order phase transition to a quark-gluon plasma phase \cite{LuciniTeper1, LuciniTeper2} with the latent heat growing as $N_c^2$.  This transition tells us nothing about a Hagedorn spectrum.  However,  systems with first order transitions can superheat and thus a hadronic phase can exist about $T_c$.   The Hagedron spectrum and the  noninteracting nature of hadrons at large $N_c$  implies a maximum temperature for this superheated hadronic phase  \cite{Cohen2006, BringolzTeper}.  Moreover, it is practical to study this superheated phase in lattice QCD for moderately large $N_c$ and thus get indirect evidence for a Hagedorn spectrum.

An alternative indirect way to demonstrate a Hagedorn spectrum for large $N_c$ QCD for 3+1 dimensions was outlined in ref.~\cite{Cohen1}.  The argument relies only standard and generally accepted properties of QCD.
Confinement in its basic sense that all physical states are color
singlets plays a critical role as does
asymptotic freedom.  In addition, the approach requires some plausible assumptions about the validity of perturbation theory to describe the correlation functions at short times.  However, the approach
explicitly assumes neither that the hadron dynamics is stringy in nature nor that the confinement
is manifest through an unbroken center symmetry.
The argument relies critically  on the fact that the number of independent local operators
with given set of quantum numbers grows exponentially with the mass dimension of operators.  This approach is similar in spirit to the ideas of Kogan and Zhitnitski \cite{KoganZhitnitsky};
it also has elements which are reminiscent of refs.~\cite{Sundborg, AharonyMarsano1, AharonyMarsano2}.

The principal purpose of this paper is to generalize the argument of ref.~\cite{Cohen1}.
The version of the argument in ref.~\cite{Cohen1} does not apply to 2+1 dimensions; here we develop
a variant of the argument which is applicable to both 3+1 and 2+1 cases.  We also simplify and clarify the arguments in  ref.~\cite{Cohen1} and improve it in significant ways.  The principal improvement is in the treatment of perturbative corrections to correlations in section~\ref{pc} of this work.  In ref.~\cite{Cohen1} corrections due to certain classes of diagram were shown to have a required behavior and it was suggested that the general case ought to behave similarly. Here a complete demonstration that this is true is given.

\section{Outline of procedure\label{proc}}

We start by noting that at large $N_c$ meson widths go to zero.  Thus the spectrum of mesons is unambiguously defined.  To begin the analysis we define  two functions characterizing the spectrum of hadrons.
$N(m)$  is defined as the number of hadrons with mass less than $m$, and $W(m)$ defined as the
sum of the masses of respective particles
\begin{equation}
W(m) = \sum\limits_{i}^{N_m} m_i = m  N(m) - \int\limits_{0}^{m}  {\rm d}\mu N(\mu) \; .
\label{functionWdef}
\end{equation}
It is easy to see that if one of them grows exponentially so does the other.

Next, we explicitly construct a sequence of sets of local operators with fixed mass dimension
(labeled with $n$) which
grows exponentially. The number of operators in each set of the sequence is
\begin{equation}
N=A^n.
\label{Nthstep}
\end{equation}
Our goal will be to demonstrate that at sufficiently large $n$ the following inequalities hold
\begin{eqnarray}
N  ( a  n + b) & \geq  & V \;; \label{conditionleft} \\
V & \geq &  W(m_N) \;, \label{conditionright}
\end{eqnarray}
where $V$ is the negative logarithmic derivative of the trace of a certain matrix of correlators and $a$ and $b$ are constants with dimensions of mass.
The key feature is that the left-hand side grows linearly with $n$ and no faster.

If we now assume that the number of hadrons  is bounded from above $N(m) \leq \exp (\alpha m)$ we can easily derive
from eqs.~(\ref{functionWdef})-(\ref{conditionright}) following expression:
\begin{equation}
a  \alpha  \log_A(e)  m + b \geq  m - \frac{1}{\alpha}  \;.
\end{equation}
As $m \rightarrow \infty$ there is a contradiction unless $\alpha \geq \frac{1}{a \log_A(e)}$ and the assumption that
function $N(m)$ is bound by exponential is false. Consistency requires
\begin{equation}
N(m) \geq \exp \left( \frac{1}{a\log_A(e)} m \right)  \;.
\end{equation}
Consequently we obtain the Hagedorn spectrum.

Inequalities (\ref{conditionleft}) and (\ref{conditionright}) are somewhat subtle and their derivation will be described in
detail in the following sections.

\section{Sets of local operators}

To proceed further we need to construct sets of composite  single-color-trace color-singlet local operators.  The matrix of correlators of operators in these sets is the core of the argument.  The single  color trace nature of these operators will ensure that at large $N_c$ each of these operators when acting on the vacuum makes a single hadron (provided that one assumes confinement) \cite{tHooft}.   These operators need to have the property that at large $N_c$, the correlator between two distinct operators in the set must vanish  as the distance between the operators goes to zero.  For simplicity we consider operators which transform as Lorentz scalars; these are guaranteed to produce spinless hadrons when acting on the vacuum (for any spatial dimension).  It is sufficient to show that that the number of spinless hadrons grows exponentially to establish a Hagedorn spectrum.

The operators we use in our construction need to be different for 2+1 and 3+1 dimensional QCD.  We will construct our operators out of some basic building blocks.
In 3+1 dimensions,  these building blocks are the following two types of operators:
\begin{equation}
O_1 = {\rm const} \cdot F_{\mu\nu}F^{\mu\nu} \,\,,\,\,\,\, O_2 ={\rm const} \cdot  F_{\mu\nu}\widetilde{F}^{\mu\nu} \;,
\label{operators3}
\end{equation}
and in 2+1 dimensions:
\begin{eqnarray}
O_1 &=&{\rm const} \cdot  F_{\alpha\beta}F^{\alpha\beta} \, F_{\alpha'\beta'}F^{\alpha'\beta'} \, F_{\alpha''\beta''}F^{\alpha''\beta''} \nonumber \;,\\
O_2 &=&{\rm const} \cdot  \left[ \epsilon^{\alpha\mu\nu}F_{\mu\nu} \, \epsilon^{\beta\mu'\nu'}F_{\mu'\nu'} \, \epsilon^{\gamma\mu''\nu''}F_{\mu''\nu''} \,\, \epsilon_{\alpha\beta\gamma}   \right]^2 \;, \label{operators2}
\end{eqnarray}
where the constants may be chosen for convenience and do not affect any results.  These operators are not traced over color.   Thus, in the large $N_c$ limit these become pure color adjoint operators. It is easy to
see that such operators are linearly independent---one is a scalar and one is a pseudoscalar.

From these basic building blocks, we create the individual color-singlet operators in the following way:
\begin{equation}
J_{l_1,l_2,\dots,l_n} = \bar{q} O_{l_1} O_{l_2} \dots O_{l_n} q   \;,
\label{currentsdef}
\end{equation}
where $l$s are either 1 or 2 and $n$ is the total number of the operators $O$ inserted between quarks.  We construct sets of these operators each of which has the same value of $n$.  Thus,
\begin{equation}
\begin{split}
&{\cal S}_1=\{ J_1, J_2 \} = \{\overline{q}O_1 q, \overline{q}O_2 q  \} \;, \\
&{\cal S}_2=\{ J_{11}, J_{12}, J_{21},J_{22} \}=  \{\overline{q}O_1  O_1q, \overline{q}O_1  O_2q,\cdots \} \;,\\
& {\cal S}_3=\{ J_{111}, J_{112}, J_{121}, J_{122}, J_{211},J_{212}, \cdots \} \;,\\
& \cdots \;.
\end{split}
\end{equation}
Ultimately we consider these a \emph{sequence} of sets where the $n^{\rm th}$ element of the sequence is ${\cal S}_n$.  We will then focus on the behavior as $n$ becomes large.  By construction, the number of currents in the $n^{\rm th}$ step of the sequence equals $N=2^n$.
Additionally, all currents in a given set ${\cal S}_n$ have the same (naive) mass dimension, $4n+3$ for 3+1 dimensions and
$12n+3$ for 2+1 dimensions, respectively.
To sum up, we created a sequence of sets of operators where the number of elements in each step grows exponentially
with a mass dimension.

\section{Matrix of current correlators}

Let us define a sequence of correlator matrices $\mathbf{\Pi}^{(n)}$ between two space-time points. Their matrix elements read
\begin{equation}
\Pi_{ab}^{(n)}(x-y) = \left \langle  J_a^{\dagger}(x) J_b(y)  \right \rangle  \;,
\end{equation}
where currents $J_{a,b} \in {\cal S}_n$. The dimension of such matrix is equal to the number of currents in the respective set, i.e.,  $2^n$.
To keep things clear, we will be using the following notation: matrices elements will always be written with explicit indices whereas  matrices themselves will be indicated by boldface
($\Pi_{ab} \leftrightarrow \mathbf{\Pi}$).

The large $N_c$ limit plus the assumption of confinement guarantees that every current generates only single meson states; the widths
go to  zero as $N_c$ approaches infinity.  Thus, the spectral decomposition of the correlator is given by
\begin{equation}
J_a(t,\vec{x}) |0\rangle = \sum\limits_{k} \int \frac{{\rm d}^3\vec{p}}{(2 \pi)^{3/2}} \,\,\, c_{ak}  \,\,\, \frac{1}{\sqrt{2E_k}} \,\,\, e^{i (E_k t-\vec{p}\cdot\vec{x})} \,\, |k,\vec{p}\rangle \;.
\label{current}
\end{equation}
Here, $c_{ak}$ is the amplitude that the current $a$ creates the particle $k$.

Using eq.~(\ref{current}) we can write the matrix $\mathbf{\Pi}$ (without loss of generality we can take $y=0$)
\begin{equation}
\Pi_{ab}^{(n)} (t,\vec{x}) = \sum\limits_{k} \int \frac{{\rm d}^3\vec{p}}{(2 \pi)^3} \,\, C_{ab,k}  \,\, \Delta(t,\vec{x}; m_k) \;,
\end{equation}
where $C_{ab,k}=c^*_{ak}c_{bk}$ is the matrix of coefficients and $ \Delta(t,\vec{x}; m_k)$ is the propagator for a noninteracting scalar of mass $m_k$ . Our matrix can be viewed as the
Kallen-Lehmann spectral representation with the spectral function $\rho$ proportional to Dirac delta functions. It is a
straightforward consequence of the large $N_c$ \cite{tHooft, Witten} limit, planarity of diagrams, and confinement.

We will study only a correlation in time ($\vec{x}=0$) and perform an analytic continuation to an imaginary time ($\tau=it$),
thus
\begin{equation}
\Pi_{ab}^{(n)}(\tau) = \sum\limits_{k} \int \frac{{\rm d}^3\vec{p}}{(2 \pi)^3} \,\, C_{ab,k} \,\,  \,\, \Delta(\tau; m_k) \;.
\label{PIabEucl}
\end{equation}

\section{The relation between matrix of correlators and masses of mesons}

In this section we derive the eq.~(\ref{conditionright}) which is at the heart of the demonstration of a Hagedorn spectrum.  We note that a derivation of this was given in ref.~\cite{Cohen1}.  Unfortunately that derivation contained an error.  The result, however, is correct and a valid derivation is given here.

Before proceeding, it is useful to recall that it is standard to use current-current correlation functions in lattice QCD to extract the lattice hadronic state in a given channel \cite{Lattice}.   The following relations are used in this context:
\begin{eqnarray}
\lim\limits_{\tau \rightarrow \infty} -\frac{\rm d}{{\rm d} \tau} \log \left \langle J(\tau)J(0) \right \rangle &=& m_0   \;,\\
 -\frac{\rm d}{{\rm d} \tau} \log \left \langle J(\tau)J(0) \right \rangle  &>& m_0  \;, \label{corin}
\end{eqnarray}
where $m_0$ is the lowest mass state.
The goal here is to generalize the relation in eq.~(\ref{corin}) to the case of a matrix of correlators $\mathbf{\Pi}^{(n)}$ in large $N_c$ QCD.

From eq.~(\ref{PIabEucl}), the matrix elements are
\begin{equation}
\Pi_{ab}^{(n)}(\tau) = \left \langle  J_a^{\dagger}(\tau) J_b(0)  \right \rangle = \sum\limits_{k} \int \frac{{\rm d}^3\vec{p}}{(2 \pi)^3} \,\, C_{ab,k} \,\,  \frac{1}{2E_k} \,\,e^{-E_k \tau}  \;,
\end{equation}
where $a,b=1 \dots 2^n$. We  study the following expression and will prove that it is greater than the sum of the lowest $2^n$  masses with scalar quantum numbers.
\begin{equation}
V^{(n)}  \equiv - \frac{\rm d}{ {\rm d} \tau }   {\rm Tr} \log \mathbf{\Pi}^{(n)}   = {\rm Tr} \left(  -\dot{\mathbf{\Pi}}^{(n)}    {\mathbf{\Pi}^{(n)}}^{-1}  \right)  \;,
\label{Vfunc}
\end{equation}
where the dot stands for the derivative with respect to Euclidean time $\tau$.

First, let us define the following quantity:
\begin{equation}
-\check{\Pi}_{ab}^{(n)}(\tau) \equiv  \sum\limits_{k} \int \frac{{\rm d}^3\vec{p}}{(2 \pi)^3} \,\, C_{ab,k} \,\,  \frac{1}{2E_k} \,\,e^{-E_k \tau} \,\, m_k  \;,
\label{hookedmatrix}
\end{equation}
compared to the derivative with respect to $\tau$:
\begin{equation}
-\dot{\Pi}_{ab}^{(n)}(\tau) = \sum\limits_{k} \int \frac{{\rm d}^3\vec{p}}{(2 \pi)^3} \,\, C_{ab,k} \,\,  \frac{1}{2E_k} \,\,e^{-E_k \tau} \,\, \sqrt{p^2+m_k^2}  \;,
\end{equation}
One can easily prove that the trace (\ref{Vfunc}) with $\check{\mathbf{\Pi}}$ instead of $\dot{\mathbf{\Pi}}$ is always
smaller than the trace with the dotted matrix:
\begin{eqnarray}
{\rm Tr} \left(  -\dot{\mathbf{\Pi}}   \right) \mathbf{\Pi}^{-1} & \geq & {\rm Tr} \left(  -\check{\mathbf{\Pi}}   \right) \mathbf{\Pi}^{-1}  \;,   \nonumber\\
\sum\limits_{c}  \langle \psi_c| -\dot{\mathbf{\Pi}} |\psi_c \rangle \lambda_c^{-1} & \geq &  \sum\limits_{c}  \langle \psi_c| -\check{\mathbf{\Pi}} |\psi_c \rangle \lambda_c^{-1} \label{ineq1} \;,
\end{eqnarray}
where $|\psi_c\rangle$ are eigenvectors of matrix $\mathbf{\Pi}$, and $\lambda_c$ are corresponding eigenvalues.
All eigenvalues, $\lambda_c$, are positive since $\mathbf{\Pi}$ is a positive definite matrix. The left part of inequality (\ref{ineq1}) represents the average
of energy (which includes momenta) in certain states, which is always greater than an average of the analogous
masses, that correspond to the term on the right-hand side.
The average is taken with positive weights $\lambda_c^{-1}$ so the inequality holds overall.

Next, we split the matrix $\mathbf{\Pi}$ into two parts, $\mathbf{A}$ and $\mathbf{B}$, where $\mathbf{A}$ couples only to the first $2^n$ masses and $\mathbf{B}$ to the rest.
\begin{eqnarray}
A_{ab}^{(n)}(\tau) &\equiv& \sum\limits_{k=1}^{2^n} \int \frac{{\rm d}^3\vec{p}}{(2 \pi)^3} \,\, C_{ab,k} \,\,  \frac{1}{2E_k} \,\,e^{-E_k \tau}  \;, \nonumber\\
B_{ab}^{(n)}(\tau) &\equiv& \sum\limits_{k=2^n+1}^{\infty} \int \frac{{\rm d}^3\vec{p}}{(2 \pi)^3} \,\, C_{ab,k} \,\,  \frac{1}{2E_k} \,\,e^{-E_k \tau} \;.  \label{AandB}
\end{eqnarray}
One can easily see that such splitting can be done for the matrix itself, $\mathbf{\Pi}=\mathbf{A}+\mathbf{B}$, its derivative, $\dot{\mathbf{\Pi}}=\dot{\mathbf{A}}+ \dot{\mathbf{B}}$, as well as
for the matrix, $\check{\mathbf{\Pi}}=\check{\mathbf{A}}+\check{\mathbf{B}}$.

Our goal is to show that we can relate the $\mathbf{\Pi}$ matrix to the meson masses.
We  start with the following simple matrix identity (which is valid for every $n$):
\begin{equation}
\begin{split}
{\rm Tr}\left ( \left( -\check{\mathbf{A}} - \check{\mathbf{B}}  \right)   \left( \mathbf{A} + \mathbf{B} \right)^{-1} \right ) & = {\rm Tr} \left( \frac{1}{\sqrt{\mathbf{A}+\mathbf{B}}} \sqrt{\mathbf{A}} \left( \frac{1}{\sqrt{\mathbf{A}}} (-\check{\mathbf{A}}) \frac{1}{\sqrt{\mathbf{A}}} \right) \sqrt{\mathbf{A}} \frac{1}{\sqrt{\mathbf{A}+\mathbf{B}}} \right)\\
& + {\rm Tr} \left( \frac{1}{\sqrt{\mathbf{A}+\mathbf{B}}} \sqrt{\mathbf{B}} \left( \frac{1}{\sqrt{\mathbf{B}}} (-\check{\mathbf{B}}) \frac{1}{\sqrt{\mathbf{B}}} \right) \sqrt{\mathbf{B}} \frac{1}{\sqrt{\mathbf{A}+\mathbf{B}}} \right) \label{mf}
\end{split}
\end{equation}
Next we introduce the definitions
\begin{eqnarray}
\mathbf{X}   \equiv \sqrt{\mathbf{A}} \frac{1}{\sqrt{\mathbf{A}+\mathbf{B}}} \,\,\,,\,\,\,\,\,\,\,\,\,\  \mathbf{Y} \equiv \sqrt{\mathbf{B}} \frac{1}{\sqrt{\mathbf{A}+\mathbf{B}}} \,\,\,,  \nonumber \\
\underline{\mathbf{A}}  \equiv   \frac{1}{\sqrt{\mathbf{A}}} (-\check{\mathbf{A}}) \frac{1}{\sqrt{\mathbf{A}}} \,\,\,,\,\,\,\,\,\,\,\,\  \underline{\mathbf{B}} \equiv  \frac{1}{\sqrt{\mathbf{B}}} (-\check{\mathbf{B}}) \frac{1}{\sqrt{\mathbf{B}}}   \;. \label{ABunderline}
\end{eqnarray}
which allows us to recast eq.~(\ref{mf}) in the following way:
\begin{equation}
 {\rm Tr}\left ( \left( -\check{\mathbf{A}} - \check{\mathbf{B}}  \right)   \left( \mathbf{A} + \mathbf{B} \right)^{-1} \right )= {\rm Tr} \left( \mathbf{X}^{\dagger} \underline{\mathbf{A}} \mathbf{X} + \mathbf{Y}^{\dagger} \underline{\mathbf{B}} \mathbf{Y} \right)  \; .
\end{equation}
The relation between $\mathbf{X}$ and $\mathbf{Y}$ operators $\mathbf{X}^{\dagger} \mathbf{X} + \mathbf{Y}^{\dagger}\mathbf{Y} = 1$ guarantees that they can be simultaneously
diagonalized and their eigenvalues are related;
\begin{equation}
\mathbf{X}^{\dagger}\mathbf{X} |\psi_i \rangle = p_i |\psi_i \rangle \,\,,\,\,\,\,\, \mathbf{Y}^{\dagger}\mathbf{Y} |\psi_i \rangle = (1- p_i) |\psi_i \rangle  \;,
\end{equation}
with  $0 \le p_i \le 1$.
We can see that the states defined as $|\chi_i \rangle \equiv 1/\sqrt{p_i} \, X|\psi_i \rangle$ forms an orthonormal basis of states, as do the states  $|\Upsilon_i \rangle \equiv1/\sqrt{1-p_i} \, Y|\psi_i \rangle$.   Thus, we can rewrite the trace into the following form.
\begin{eqnarray}
{\rm Tr} \left( \mathbf{X}^{\dagger} \underline{\mathbf{A}} \mathbf{X} + \mathbf{Y}^{\dagger} \underline{\mathbf{B}} \mathbf{Y} \right) &=& \sum\limits_{i} \langle \psi_i| \mathbf{X}^{\dagger} \underline{\mathbf{A}} \mathbf{X} + \mathbf{Y}^{\dagger} \underline{\mathbf{B}} \mathbf{Y} |\psi_i \rangle \nonumber \\
&=&  \sum\limits_{i} \langle \chi_i| \underline{\mathbf{A}} |\chi_i \rangle + (1-p_i) \left(  \langle \Upsilon_i| \underline{\mathbf{B}} |\Upsilon_i \rangle -  \langle \chi_i| \underline{\mathbf{A}} |\chi_i \rangle      \right) \label{trace1}\\
&\geq& {\rm Tr} \underline{\mathbf{A}} = {\rm Tr} \frac{-\check{\mathbf{A}}}{\mathbf{A}}  \;.
\end{eqnarray}
The second term of the right-hand side of eq.~(\ref{trace1}) is always positive since the biggest possible eigenvalue of
operator $\underline{\mathbf{A}}$ is less than or equal to $m_{2^n}$ (where $m_j$ is the mass of the $j^{\rm th}$ state)  , while the smallest
eigenvalue of $\underline{\mathbf{B}}$ is always greater or equal to $m_{2^n+1}$.  This is a straightforward
consequence of eq.~(\ref{hookedmatrix}), (\ref{AandB}), and (\ref{ABunderline}).

The currents entering operator $\mathbf{A}$ can be reorganized in such a way  that the first current couples only to the first meson, second current
only to the second meson, etc. It is obvious that this process will lead to the diagonal matrix whose every element is equal to the mass
of the $n^{\rm th}$ particle. Consequently
\begin{equation}
{\rm Tr} \frac{-\check{\mathbf{A}}^{(n)}}{\mathbf{A}^{(n)}} = \sum\limits_{k}^{2^n} m_k  \;.
\end{equation}

The quantity on the right-hand side is exactly the function $W(m_{2^n})$ defined by eq.~(\ref{functionWdef}). Consequently,
we derived the inequality of (\ref{conditionright}) required for the establishing of a Hagedorn spectrum.
\begin{equation}
V= - \frac{\rm d}{ {\rm d} \tau }   {\rm Tr} \log \mathbf{\Pi}     \geq  \sum\limits_{k}^{2^n} m_k = W(m_{2^n})  \;.
\end{equation}

\section{Matrix of correlators in an asymptotically free regime\label{AF}}

In this section, we derive the condition (\ref{conditionleft}) stating that the trace of
the logarithm of a corellator matrix grows at most linear in $n$ (where $n$ labels the step in the sequence).

For sufficiently small times, asymptotic freedom allows us to treat fields inside the currents as
non-interacting, so we can decompose the correlator to a product of
single-particle propagator functions. Additionally, the large $N_c$ limit, in which we are working,
guarantees that the whole matrix of correlators is diagonal. Doing the trace, we obtain $2^n$ terms with the same structure.
Consequently, we can focus solely on the one current-current correlator and its logarithmic derivative, and
investigate how it grows with $n$.

In the asymptotically free region, the structure of the overall correlator is given simply by the dimensional analysis.
Since the time scale $\tau$ is the only dimensional parameter left, and our current $J$ has the mass
dimension $4n+3$ for 3+1 dimensions, and $12n+3$ for 2+1 dimensions, the right-hand side equals $8n+6$, and $24n+6$, respectively.
Thus
\begin{equation}
\Pi_{ab}^{(n)} = \left\{ \begin{array}{lcl}
 \delta_{ab} \, {\rm const} \,\, \tau^{-8n+6} & \mbox{for} & 3+1 \,{\rm dimensions} \\
 \delta_{ab} \, {\rm const} \,\, \tau^{-24n+6}  & \mbox{for} & 2+1 \,{\rm dimensions}
\end{array}\right   .
\label{Pin}
\end{equation}
and the trace of logarithmic derivative equals
\begin{equation}
- \frac{\rm d}{ {\rm d} \tau }   {\rm Tr} \log {\mathbf{\Pi}^{(n)}} = \left\{ \begin{array}{lcl}
 (2^n) \frac{8n+6}{\tau} & \mbox{for} & 3+1 \,{\rm dimensions} \\
 (2^n)  \frac{24n+6}{\tau}  & \mbox{for} & 2+1 \,{\rm dimensions}
\end{array}\right   .
\label{Pin2}
\end{equation}
If we assume that the matrix of correlators is effectively at its asymptotic value up to some small corrections
for $\tau<\tau_0$ with $\tau_0$ independent of $n$, we reproduced exactly the inequality condition
(\ref{conditionleft}) necessary for establishing the Hagedorn spectrum.

From the argument in section~\ref{proc}, the preceding implies a  Hagedorn spectrum where the value of Hagedorn temperature corresponding to our sets of currents is
\begin{equation}
T_H \leq \left\{ \begin{array}{lcl}
 \frac{8 \log_2(e)}{\tau_0} & \mbox{for} & 3+1 \,{\rm dimensions} \\
 \frac{24 \log_2(e)}{\tau_0}  & \mbox{for} & 2+1 \,{\rm dimensions}
\end{array}\right   .
\end{equation}

At this stage, we have shown that a Hagedorn spectrum emerges in the QCD in 3+1 and 2+1 dimensions if a
certain assumption is met---namely that correlators are to good approximation at their
asymptotically free value for $\tau < \tau_0$ for all $n$.

\section{Perturbative corrections\label{pc}}

In the previous section, we neglected all possible interactions between gluons. Such  an assumption
was justified by the fact that the QCD is in asymptotically free regime for short times. Provided
that this condition is met, we have proved QCD has a Hagedorn spectrum.  The critical question is then the circumstances for which
there are no large corrections to the asymptotically free result.

The standard way  to include the effects of  interactions for the correlators at short times  is via perturbation theory.
However, the region where perturbation theory is valid is certainly limited; as the time increases
perturbative corrections grow and ultimately push the system outside the region of validity of perturbation theory.
Here we will rely  on the standard assumption that perturbation theory accurately describes correlation functions provided that they are small.  That is, in the region where pertrubative corrections are small they will dominate over all nonperturbative effects.  We note that this is not a rigorous mathematical theorem but it is the basis of  standard analysis of QCD correlation functions.

Given this assumption, the critical issue we need to address is how do perturbative corrections  scale with $n$?
The goal is to show that at fixed $\tau$ perturbative corrections, at any fixed order, to the quantity
$2^{-n}\frac{\rm d}{ {\rm d} \tau }{\rm Tr} \log {\Pi^{(n)}} $ scales at most linearly with $n$ for any fixed $\tau$ (in order to satisfy
the inequality~(\ref{conditionleft})).  If we can demonstrate this, we have demonstrated a Hagedorn spectrum given the assumptions stated above.  To see why,  imagine doing the following calculation:  start at some fixed but large $n$ and  compute
$2^{-n}\frac{\rm d}{ {\rm d} \tau }{\rm Tr} \log {\Pi^{(n)}} $ in perturbation theory to some order.  Next decrease the value of $\tau$ so that perturbative corrections are sufficiently small that the quantity is close to its asymptotically free value,  up to corrections which are a small fraction of the total.  It is always possible to find a value of $\tau$ for which this is true since the system  becomes asymptotically  free as $\tau \rightarrow 0$.    By assumption this is the regime in which perturbation theory is trustworthy.   Having done this  at some fixed value of $n$,  one next increases $n$ keeping $\tau$ fixed.  Since the perturbative corrections to
$2^{-n}\frac{\rm d}{ {\rm d} \tau }{\rm Tr} \log {\Pi^{(n)}} $  have been demonstrated to scale at most as $n$ and since, as seen in eq.~(\ref{Pin2}), the leading behavior from the asymptotically free region also scales with $n$, the fraction size of the correction is independent of $n$.  Since the corrections to the asymptotically free result were small at the original $n$, they remain small at all $n$ including in the limit of $n \rightarrow \infty$.  This is sufficient to show a Hagedorn spectrum, given the assumptions stated above, and given the result of section~\ref{AF}.

In the remainder of this section we show that perturbative corrections to \\
$2^{-n}\frac{\rm d}{ {\rm d} \tau }{\rm Tr} \log {\Pi^{(n)}} $ do,
in fact, scale with $n$ no faster than linearly and thereby complete the demonstration.   We note that an argument that this is the case was presented in ref.~\cite{Cohen1}.  In that work, the results of certain classes of diagrams were presented and shown to be consistent with the needed result.  It is was suggested that the structure of the quantity ought to ensure that result continued to hold for all classes of diagram, however, no demonstration of that was given.  Here we construct a general argument why the result will hold for all classes of diagram.

\begin{figure}
\begin{center}

\includegraphics[width=7cm]{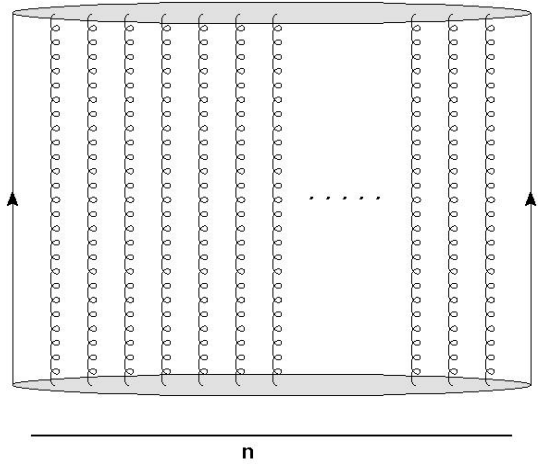}

\caption{The Feynman diagram for a correlator in the asymptotically free regime.  The blobs indicate the currents.}

\label{feyn_diag_free}

\end{center}
\end{figure}

In order to get more insight into the correlation functions, we first  look at the contribution from non-interacting gluons such as in the  Feynman diagram
illustrated in figure~\ref{feyn_diag_free}.
It will allow us to extract some properties that we will generalize later when we include interactions.
In the non-interacting case, the correlator matrix is diagonal due to large $N_c$ limit.
Each propagating gluonic operator  $O_l$ inside the current contributes
the same way
\begin{equation}
\Pi^1_{i j \, \rm free}(\tau) = \left\langle O_i(\tau) \, O_j(0) \right\rangle_{\rm free} = \delta_{ij} \, \pi^1_{\rm free}(\tau)    \;,
\end{equation}
where the superscript $1$ indicates that this represents the propagation of a single gluonic operator.   The $\delta_{i j}$ is a result of the choice of operators (\ref{operators3}), (\ref{operators2}) which were picked precisely because of this property.  The fact that the free correlator  is the same for both operators is a result of dimensional analysis which requires them to be proportional to each other; a choice of the constants defining the operators can fix the proportionality constant to unity.
We want to emphasize that this object is not gauge invariant, however, it will be only a part of the overall correlator
which will be gauge invariant.
The small letter $\pi$ indicates that the quantity is already a function, not a matrix. Such convention will be applied also in the following text.
Let us also remind the reader that the matrix elements explicitly indicated indices, whereas the analogous matrices
are denoted in boldface.

The correlation function of the whole current consists of $n$ internal lines and is bounded
by two quark lines.  Thus its matrix elements are given by
\begin{equation}
\Pi^{(n)}_{a b}(\tau)  = \delta_{a b} \; \pi^q_{\rm free}(\tau) \left[ \pi^1_{\rm free} (\tau) \right]^n  \pi^q_{\rm free}(\tau) \;.
\end{equation}
where $  \pi^q_{\rm free}(\tau) $ is the free quark propagator traced over Dirac indices.
 Recall that the indexes $a$, $b$ goes from $1$ to $2^n$.  The quantity of interest  is the derivative of trace of the logarithm,
\begin{equation}
2^{-n} \frac{\rm d}{{\rm d}\tau} {\rm Tr} \, \log {\mathbf{\Pi}^{(n)}} = 2 \frac{\rm d}{{\rm d}\tau}\log \pi^q_{\rm free} + n \frac{\rm d}{{\rm d}\tau} \log \pi^1_{\rm free}  \;.
 \label{tracelogfree}
\end{equation}
The first term is independent of $n$ (and comes for 2 quarks on the boundaries), while the second one is linearly proportional to $n$.
This result is in agreement with the results obtained in the previous section.

\begin{figure}
\begin{center}

\includegraphics[width=7cm]{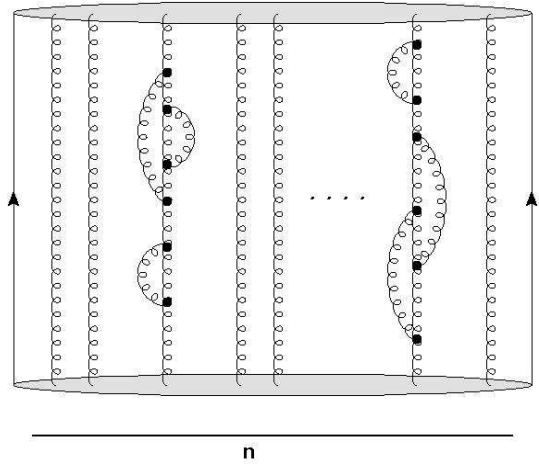}

\caption{An example of a Feynman diagram  where interactions do not couple distinct gluon lines connected to the sources.}
\label{feyn_diag_1_int}

\end{center}
\end{figure}

Diagrams with interactions are more complicated. Let us work in a case, when all interaction up to order $\alpha^l$ are included.
First, we restrict our attention to the case where all of the interactions act on single gluons  that are connected to the sources (see figure~\ref{feyn_diag_1_int}).
While more than one of these gluons may be involved, there are no interactions which couple distinct gluons coupled to the source in this class. Effectively, these interactions lead  to the modification of a free propagator
 similar to the contribution of self-energy correction.
\begin{equation}
\pi^1_{\rm free}(\tau)  \rightarrow  \pi^1_{\rm free}(\tau) \, \left( 1+c \left( \tau \right)\right)  \;,
\end{equation}
where $(1+c(\tau))$ is a pertrubative correction. Note that $c$ depends on the order to which we work in perturbation theory,
but it is well defined at any given order.
Such correction can appear on any internal gluonic line, so that if these were the only types of diagrams contributing we could write the total correlator as
\begin{equation}
\Pi^{(n)}_{ab}(\tau) = \delta_{ab} \; \pi^q_{\rm free}(\tau)    \left[ \pi^1_{\rm free}(\tau) \, (1+c(\tau) ) \right]^n   \pi^q_{\rm free}(\tau)  \; .
\end{equation}
This structure actually contains more information (higher order in $\alpha$) than necessary, but certainly
contains all combinations that are required to order $\alpha^l$.
The trace of the logarithm now reads
\begin{equation}
 2^{-n} \frac{\rm d}{{\rm d}\tau} {\rm Tr} \, \log {\mathbf{\Pi}^{(n)}} = 2 \frac{\rm d}{{\rm d}\tau}\log \pi^q_{\rm free} + n \frac{\rm d}{{\rm d}\tau} \log \pi^1_{\rm free} + n \frac{\rm d}{{\rm d}\tau} \log (1+c)  \;.
 \label{tracelogone}
\end{equation}
Although we have yet to calculate $c$, it is clearly independent of $n$. So the total expression grows, again, at most linearly with $n$.  Of course, this result is wrong---the class of diagrams we considered was chosen artificially and does not correspond to the full perturbative result at any order in $\alpha$.  However, it does illustrate how the  logarithmic structure combined with factorizing point-to-point correlators for the individual lines yields   total perturbative corrections which grow at most with $n$.
Our main task is to show that a structure enriched with inter-gluonic interactions obeys the same at most linear $n$ dependant growth.

\begin{figure}
\begin{center}

\includegraphics[width=7cm]{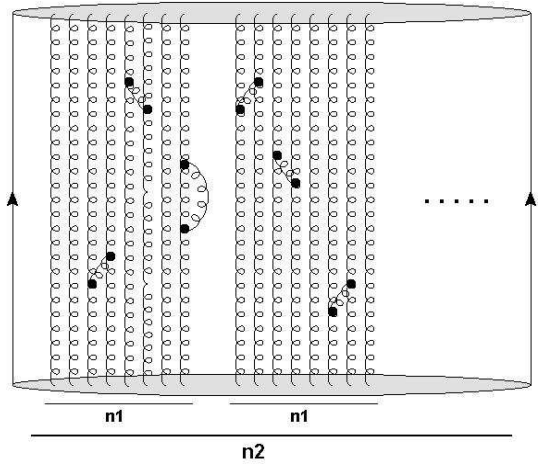}

\caption{Division of gluon lines into $n_2$ clusters each containing $n_1$ lines.}
\label{feyn_diag_8}

\end{center}
\end{figure}

Up to now, we have ignored possible interactions between gluons.
In order to deal with them in a simple way, we artificially divide the total number of internal gluon lines $n$ into $n_2$ clusters each
containing $n_1$ gluon lines, that is $n=n_1  n_2$. We will impose the conditions $n_1\gg 1$, $n_2\gg 1$; and $n_1 \gg$~the order
in perturbation theory to which we are working.
For the beginning, let us assume that all interactions occur within clusters,
as is illustrated in  the figure~\ref{feyn_diag_8}, i.e., there is no internal line between clusters or between these clusters with quark lines bounding them.
Obviously, in doing this we neglect a certain class of diagrams for now.   However, we will subsequently show that correction due to their inclusion does not affect the leading $n$ behavior.   The reason for this is that for a fixed order of perturbation theory only a very small fraction of all Feynman diagrams will connect different clusters and this fraction is small enough to alter leading behavior.

The propagation within  one cluster is given by
\begin{equation}
\mathbf{\Pi}^C(\tau)= \left[ \pi^1_{\rm free}(\tau) \right]^{n_1} \left[ \mathbf{1}+\mathbf{C}'(\tau,n_1) \right]   \;,
\end{equation}
where $\left[\mathbf{1}+ \mathbf{C}'(\tau , n_1)\right]$ represents the effect of interactions within one cluster.
Note that $\mathbf{\Pi}^C$ is a matrix of the dimension  $2^{n_1}\times 2^{n_1}$.
The $\mathbf{C}'(\tau, n_1)$ depends on the order to which we work in perturbation theory, and, obviously, it depends
on the size of the cluster, the number of internal lines $n_1$.

Now, we need to define a mapping from the ``cluster''space,
which has the dimension $2^{n_1}$ matrix to the ``overall correlator'' space with the dimension $2^n$, $\mathbf{D}(\mathbf{M})$.
More precisely, we actually need  $n_2$ different mappings $\mathbf{D}^{(k)}(\mathbf{M})$. each corresponding to a different cluster.

Our mapping will be  a $2^n \times 2^n$ matrix $D^{(k)}_{ab}$. Note that indexes $a$ and $b$ can be represented
by a sequence of $n$ numbers, $a=(a_1a_2\dots a_n)$, $b=(b_1b_2\dots b_n)$,
 where $a_i,b_i=0,1$. It is a straightforward consequence of our
original definition of currents (\ref{currentsdef}), each of them being constructed from $n$ building block operators
(translated into $n$ internal gluonic lines in each diagram).
Recall that we divided $n$ lines into $n_2$ clusters of $n_1$ elements .  Moreover since we neglect interactions {\it between} clusters, we want to treat them independently.  Thus it is useful to
 define a mapping
$\mathbf{D}^{(k)}$ to work in such a way  that the first one, $\mathbf{D}^{(1)}$, affects only first $n_1$
subindices within  indices $a$ and $b$, the mapping corresponding to the second one, $\mathbf{D}^{(2)}$, affects pieces $n_1+1$ to $2n_1$, etc.
Analogously, all $n_2$ mappings corresponding to all possible $n_2$ clusters are defined.
Specifically,
\begin{eqnarray}
D^{(1)}_{(a_1a_2\dots a_n)(b_1b_2\dots b_n)} (\mathbf{M}) &=& M_{(a_1\dots a_{n_1})(b_1\dots b_{n_1})} \;\prod_{l=n_1+1}^n \delta_{a_lb_l} \;, \nonumber\\
D^{(2)}_{(a_1a_2\dots a_n)(b_1b_2\dots b_n)} (\mathbf{M}) &=& M_{(a_{n_1+1}\dots a_{2n_1})(b_{n_1+1}\dots b_{2n_1})} \; \prod_{l=1}^{n_1} \delta_{a_lb_l} \;  \prod_{l'=2n_1+1}^{n} \delta_{a_{l'}b_{l'}} \;, \nonumber
\end{eqnarray}
and the general form
\begin{equation}
D^{(k)}_{(a_1a_2\dots a_n)(b_1b_2\dots b_n)} (\mathbf{M}) = M_{(a_{(k-1)n_1+1}\dots a_{kn_1})(b_{(k-1)n_1+1}\dots b_{kn_1})}
\; \prod_{l=1}^{(k-1)n_1} \delta_{a_lb_l} \; \prod_{l'=kn_1+1}^{n} \delta_{a_{l'}b_{l'}} \;,
\end{equation}
where $\mathbf{M}$ is an $n_1 \times n_1$ matrix.
Note that such matrices are a straightforward generalization of the previous  case where the equivalent of
$\mathbf{\Pi^C}$ was just number (matrix $1\times 1$)  and the matrix $\mathbf{D}$ was diagonal.
It is worth mentioning that the $2^n$ dimensional overall space can be factorized as a product of $n_2$ subspaces
with dimensions $2^{n_1}$ with each subspace corresponding to one particular cluster.
So, the vectors have the form
\begin{equation}
|V\rangle = |v^{(1)}\rangle \otimes |v^{(2)}\rangle \otimes \dots \otimes |v^{(n_2)}\rangle \;,
\end{equation}
and the spirit of the mapping $\mathbf{D}^{(k)}$ is
\begin{equation}
\mathbf{D}^{(k)}(\mathbf{M}) = \mathbf{1}^{(1)} \otimes \mathbf{1}^{(2)} \otimes \dots \otimes \mathbf{M}^{(k)} \otimes \dots \otimes \mathbf{1}^{(n_2)} \; .
\end{equation}
From the construction of $\mathbf{D}^{(k)}(\mathbf{M})$ it is easy to show that ${\rm Tr} \, \log \mathbf{D}^{(k)}(\mathbf{M}) ={\rm Tr} \, \log \mathbf{M}$.

Using this notation and imposing the condition that we neglect all diagrams connecting clusters, the overall correlator contains a product of $n_2$ matrices $\mathbf{D}^{(k)}$ corresponding to the respective clusters
\begin{equation}
\mathbf{\Pi}^{(n)} = \pi^q_{\rm free}(\tau) \;\; \mathbf{D}^{(1)}(\mathbf{\Pi}^C) \;\;\times\;\; \mathbf{D}^{(2)}(\mathbf{\Pi}^C)  \;\; \dots \;\; \mathbf{D}^{(n_2)}(\mathbf{\Pi}^C) \;\;  \pi^q_{\rm free}(\tau)  \; .
\end{equation}
Using the general property that ${\rm Tr} \, \log (\mathbf{A B})={\rm Tr} \, \log \mathbf{A}+ {\rm Tr} \, \log \mathbf{B}$ and the property that ${\rm Tr} \, \log \mathbf{D}^{(k)}(\mathbf{M}) ={\rm Tr} \, \log \mathbf{M}$, it is straightfoward to show that
\begin{equation}
  2^{-n} \frac{\rm d}{{\rm d}\tau} {\rm Tr} \, \log {\mathbf{\Pi}^{(n)}} = 2 \frac{\rm d}{{\rm d}\tau}  \log \pi^q_{\rm free} + n_1n_2 \frac{\rm d}{{\rm d}\tau} \log \pi^1_{\rm free} + n_2 \frac{\rm d}{{\rm d}\tau} \log (1+c'(n_1))
\label{tracelogalmostfull1}
\end{equation}
{\rm where} $ c'(n_1) \equiv \exp\left( {\rm Tr} \, \log  \left ( \mathbf{1} + \mathbf{C'}   \right) \right ) -1$.

The first two terms were already discussed after   eq.~(\ref{tracelogfree}).  The second term becomes obvious once one recalls that   $n_1n_2=n$.
The third term requires  more care.
We can  denote the expression $\frac{\rm d}{{\rm d}\tau} \log (1+c'(n_1))$ as  $f(n_1)$ , i.e., some at present unknown function of $n_1$.  However,  our choice of clusters is completely arbitrary.  Provided  our assumption that diagrams connecting clusters does not affect the leading behavior is correct,
we can switch  $n_1$ and $n_2$ and the result must remain unchanged.
This requires $n_2 f(n_1) = n_1 f(n_2)$ and consequently $f(m)$ must be a linear function of $m$.
Thus, the third term on the right-hand side of eq.~(\ref{tracelogalmostfull1}) is also proportional to $n=n_1n_2$,
just as in  eq.~(\ref{tracelogone}).

\begin{figure}
\begin{center}

\includegraphics[width=7cm]{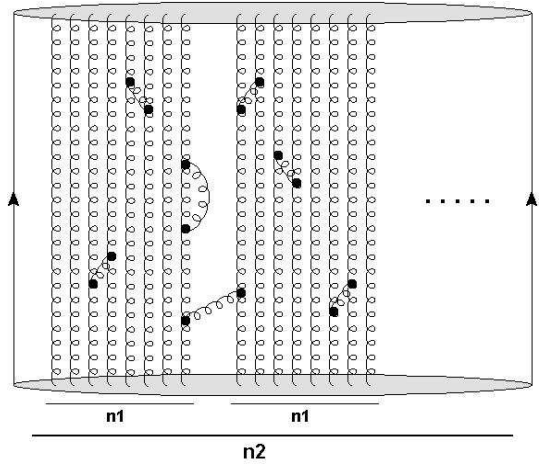}

\caption{An example of a Feynman diagram with interaction between two clusters.}
\label{feyn_diag_8_cross}

\end{center}
\end{figure}

To complete the demonstration we need to show that the inclusion of diagrams
 with interactions connecting individual clusters does not affect the leading scaling. An example
of such diagram is in  figure~\ref{feyn_diag_8_cross}.  These effects can be accounted for
 by including a matrix $(\mathbf{1}+\mathbf{C}'')$ between the matrices corresponding to clusters.
Additionally, one should include the interaction between the quarks on the boundary
and the neighboring clusters of gluons $(\mathbf{1}+\mathbf{C}^q)$.
The total correlator matrix reads
\begin{eqnarray}
\mathbf{\Pi}^{(n)}(\tau) &=& \pi^q_{\rm free}(\tau) \;\;(\mathbf{1}+\mathbf{C}^q) \;\;\times\;\; \mathbf{D}^{(1)}(\mathbf{\Pi}^C) \;\;\times\;\; (\mathbf{1}+ \mathbf{C}'')\;\;\times\;\; \mathbf{D}^{(2)}(\mathbf{\Pi}^C) \;\;\dots \nonumber\\
&& (\mathbf{1}+ \mathbf{C}'')\;\;\times\;\; \mathbf{D}^{(n_2)}(\mathbf{\Pi}^C) \;\;\times\;\; \mathbf{D}^{(1)}(\mathbf{\Pi}^C) \;\; (\mathbf{1}+\mathbf{C}^q) \;\; \pi^q_{\rm free}(\tau)  \; .
\end{eqnarray}
The key point here is that the matrices $\mathbf{C}''$ and $\mathbf{C}^q$ must be independent of $n_1$.  The reason for this is that by construction,  $n_1$ is much larger than the order in perturbation theory to which we are working.  At large $N_c$ only planar diagrams contribute.  Thus if we are working at order $\alpha_s^l$, a diagram connecting two clusters can at most go $l-1$ gluon lines into the cluster.  Since this is less than $n_1$ it cannot go across the cluster. Thus, the dynamics in  $\mathbf{C}''$ does not know how large the cluster is and must be independent of $n_1$.  It is clearly independent of $n_2$ either.

Using an analogous argument as earlier, the trace of the logarithm reads
\begin{eqnarray}
 2^{-n} \frac{\rm d}{{\rm d}\tau} {\rm Tr} \, \log {\mathbf{\Pi}^{(n)}} &=& 2 \frac{\rm d}{{\rm d}\tau}  \log \left(\pi^q_{\rm free}(1+c^q)\right)  + n_1n_2 \frac{\rm d}{{\rm d}\tau} \log \pi^1_{\rm free} \nonumber\\
 &&+ n_2 \frac{\rm d}{{\rm d}\tau} \log (1+c'(n_1)) + (n_2-1) \frac{\rm d}{{\rm d}\tau} \log (1+c'')  \;
\label{tracelogfull}
\end{eqnarray}
where $c^q$ and $c''$ are defined analogously to $c'$.  The first term does not scale with $n$; the second term is directly proportional to $n$.  Defining $f(n_1) \equiv \frac{\rm d}{{\rm d}\tau} \log (1+c'(n_1))$ and $g \equiv \frac{\rm d}{{\rm d}\tau} \log (1+c'')$,  the last two terms can be written as $n_2 (f(n_1) +g) -g$.  Again exploiting the fact that we can switch $n_1$ and $n_2$ without affecting the result (provided the order in perturbation theory is less than both $n_1$ and $n_2$) we obtain the consistency condition that
\begin{equation}
n_2  (f(n_1) +g)=n_1  (f(n_2) +g)
\end{equation}
which yields $f(m)= b m-g$ where $b$ is a constant.  Thus, the effect of including the interactions between the clusters simply fixes the subleading behavior in $f$.  Taking the last two terms together and exploiting the fact that $n=n_1 n_2$ we have $b n -g$ which grows linearly in $n$.  Consequently, the right-hand side of eq.~(\ref{tracelogfull}) grows at most linearly with $n$, and
the inequality (\ref{conditionleft}) is satisfied if the interactions are included via perturbation theory at any fixed order.

With this we have completed our demonstration that a Hagedorn spectrum arises in large $N_c$ QCD in both 2+1 and 3+1 dimensions.  This demonstration depends on one critical assumption: that perturbation theory accurately describes the trace of the  logarithm of a matrix of  point-to-point correlation functions in the regime where the perturbative  corrections to the asymptotically free value are small.

\acknowledgments This work was supported by the U.S.~Department of Energy
through grant DE-FG02-93ER-40762.

\begin{thebibliography}{99}

\bibitem{Hagedorn1}  R. Hagedorn, {\it Statistical thermodynamics of strong interactions at high energies}, Nuovo Cimento Suppl.  {\bf 3} (1965) 147.

\bibitem{Hagedorn2}  R. Hagedorn, {\it Hadronic matter near the boiling point}, Nuovo Cimento {\bf 56A} (1968) 1027.

\bibitem{bronflorgloz}  W. Broniowski,  W. Florkowski, L.Y. Glozman, {\it Update of the Hagedorn mass spectrum}, Phys. Rev. D {\bf 70} (2004) 117503-1 [hep-ph/0407290].

\bibitem{Cohen2006} T. D. Cohen, {\it QCD strings and the thermodynamics of the metastable phase of QCD at large $N_c$}, Phys. Lett. B {\bf 637} (2006) 81 [hep-th/0602037].

\bibitem{PDG} K. Nakamura and Particle Data Group, {\it Review of Particle Physics }, J. Phys. {\bf G 37} (2010) 075021.

\bibitem{ChengChrist}  M. Cheng, N. H. Christ, P. Hegde, F. Karsch, Min Li, M. F. Lin, R. D. Mawhinney, D. Renfrew, P. Vranas, {\it The finite temperature QCD using 2+1 flavors of domain wall fermions at $N_t = 8$}, Phys. Rev. {\bf D 81}  (2010) 054510  [hep-lat/0911.3450].

\bibitem{Strings}  J. Polchinski, {\it String Theory}, Cambridge University Press (1998).

\bibitem{Lattice}  H. J. Rothe, {\it Lattice Gauge Theories}, World Scientific Publishing (2005).

\bibitem{tHooft} G. t'Hooft, {\it A planar diagram theory for strong interactions}, Nucl. Phys. {\bf B 72} (1974) 461.

\bibitem{Witten} E. Witten, {\it Baryons in the 1/N expansion}, Nucl. Phys. {\bf B 160} (1979) 57.

\bibitem{KoganZhitnitsky} I.I. Kogan, A.R. Zhitnitsky, {\it Two dimensional QCD with matter in adjoint representation:
What does it teach us?},  Nucl. Phys. {\bf B 465} (1996) 99 [hep-ph/9509322].

\bibitem{DalleydeSande}  S. Dalley, B. van de Sande, {\it Finite temperature gauge theory from the transverse lattice }, Phys. Rev. Lett. {\bf 95} (2005) 162001 [hep-ph/0409114].

\bibitem{LuciniTeper1} B. Lucini, M. Teper, U. Wenger, {\it The high temperature phase transition in SU(N) gauge
theories}, JHEP {\bf 01} (2004) 061 [hep-lat/0307017].

\bibitem{LuciniTeper2} B. Lucini, M. Teper, U. Wenger, {\it Properties of the deconfining phase transition in SU(N)
gauge theories},  JHEP {\bf 02} (2005) 033 [hep-lat/0502003].

\bibitem{BringolzTeper} B. Bringoltz, M. Teper, {\it In search of a Hagedorn transition in SU(N) lattice gauge
theories at large-N}, Phys. Rev. {\bf D 73} (2006) 014517l [hep-lat/0508021].

\bibitem{Cohen1} T. D. Cohen, {\it QCD and the Hagedorn spectrum}, JHEP {\bf 06} (2010) 098	[hep-th/0901.0494].

\bibitem{Sundborg} B. Sundborg, {\it The Hagedorn Transition, Deconfinement and N = 4 SYM Theory}, Nucl. Phys. {\bf B 573} (2000) 349 [hep-th/9908001].

\bibitem{AharonyMarsano1} O. Aharony, J. Marsano, S. Minwalla, K. Papadodimas, M. van Raamsdonk, {\it The
Hagedorn/deconfinement phase transition in weakly coupled large-N gauge theories}, Adv. Theor. Math. Phys. {\bf 8} (2004) 603 [hep-th/0310285].

\bibitem{AharonyMarsano2}  O. Aharony, J. Marsano, S. Minwalla, K. Papadodimas, M. van Raamsdonk, {\it A first
order deconfinement transition in large-N Yang-Mills theory on a small 3-sphere}, Phys. Rev. {\bf D 71} (2005) 15018 [hep-th/0502149].



\end{thebibliography}
\end{document}